# An Efficient Digital Watermarking Algorithm Based on DCT and BCH Error Correcting Code

Saeideh Nabipour, Javad Javidan, Majid Khorrami, Jila Azimzadeh
University of Mohaghegh Ardabili, Department of Electronic and Computer Engineer, Ardabil, Iran
Email: saeideh.nabipour@gmail.com

*Abstract*— Watermarking is a technique for hiding of data in a medium coverage so that its presence is not detectable by a human eye and is recoverable only by the authorized recipient. Two of the most important features of watermarked image are transparency and robustness which are largely related to the security of watermarking algorithms. In this paper, an image watermarking scheme based on BCH error correction code in Discrete Cosine Transformation (DCT) domain is considered. Before embedding process, the watermark is encoded through BCH coding. Then it is embedded into the Discrete Cosine Transformation (DCT) coefficients of cover image. In order to decrease embedding complexity and speed up the process of finding the best position to insert a watermark signal, lookup table method is utilized. The key features of proposed method include the reduction of time required in the process embedding of information, security and ability to correct the error caused by variety of attacks and destructions as well. Watermarked image robustness has been investigated against different kinds attacks and the simulation results indicate that the proposed algorithm outperforms the existing methods in terms of imperceptibility, robustness and security.

## I. Introduction

Due to the rapid development of the computer network and internet, access to information, storage, copy and modify them, even diffuse illegally unauthorized copies of copyrighted digital contents without altering their quality have become more convenient. Therefore, security of information has become an indispensable factor in information transmission. One of the most effective solutions to overcome this problem is the digital watermarking. Watermarking is a technique for hiding of data in a medium coverage so that its presence is not detectable by a human visual system and is recoverable only by the authorized recipient. According to working domain, the watermarking techniques can be divided into two types: a) spatial domain watermarking techniques b) frequency domain watermarking techniques. In spatial domain techniques, the watermark embedding is done on image pixels while in frequency domain watermarking techniques the embedding is done after taking image transforms [1].

In recent years, error detecting and correcting codes has been able to have a significant impact on the robustness, security and transparency of digital image watermarking. The theory of error detecting and correcting codes is that branch of engineering and mathematics which deals with the reliable transmission and storage of data [2]. Regarding this concept, recently error correction codes are used considerably to achieve the imperceptibility, robustness and security of digital image watermarking and various codes including Hamming code [3], BCH code [4-8], Reed Solomon code [9, 10], convolutional code [11], LDPC code [12] and turbo code [13] have been studied to achieve these objectives. Several researches have been done on error correction code in digital watermarking process. For example, Marvel [14] has considered protection of the steganographic payload in images by using of Reed-Solomon codes, turbo codes and special codes developed by Retter2 in both hard decision and soft-decision decoding contexts. The scheme consisting in an interleaver that disperses long error bursts followed by the coder unit, achieves a Bit Error Rate (BER) of about $10^{-2}$ at a high embedding density of 0.16 bit/pixel.

Alattar et al. [15] suggested a watermarking algorithm for electronic documentation which uses spread spectrum method. They used BCH coding to obtain the watermark more successfully after the print and scan process of the electronic documentation. Findik et al. [16] are suggested a digital color image watermarking technique based on artificial neural networks (ANN) and BCH coding together.

Crandall [17] has considered the Matrix Embedding (ME) scheme to reduce the number of required changes of the cover by carefully selecting the positions used for embedding. Fridrich et al. [18, 19] suggested an embedding algorithm based on syndrome coding using random codes, is called Wet Paper codes. This embedding scenario does not require the sender to

Manuscript received October 9, 2016. (Write the date on which you submitted your paper for review.) This work was supported in part by the U.S. Department of Commerce under Grant BS123456 (sponsor and financial support acknowledgment goes here). Paper titles should be written in uppercase and lowercase letters, not all uppercase. Avoid writing long formulas with subscripts in the title; short formulas that identify the elements are fine (e.g., "Nd–Fe–B"). Do not write "(Invited)" in the title. Full names of authors are preferred in the author field, but are not required. Put a space between authors' initials.

F. A. Author is with the National Institute of Standards and Technology, Boulder, CO 80305 USA (corresponding author to provide phone: 303-555-5555; fax: 303-555-5555; e-mail: author@ boulder.nist.gov).

S. B. Author, Jr., was with Rice University, Houston, TX 77005 USA. He is now with the Department of Physics, Colorado State University, Fort Collins, CO 80523 USA (e-mail: author@lamar. colostate.edu).

T. C. Author is with the Electrical Engineering Department, University of Colorado, Boulder, CO 80309 USA, on leave from the National Research Institute for Metals, Tsukuba, Japan (e-mail: author@nrim.go.jp).

share any knowledge about the constraints with the recipient and does not even sacrifice embedding capacity, however, this happens at the cost of an increased embedding complexity.

Westfeld [20] suggested to use a matrix encoding technique for hiding data to DCT coefficients. His scheme hides more than one bit by changing at most one coefficient in a block. The matrix encoding technique is based on the Hamming code. The (*n, m, t*) matrix encoding technique can hide *n* bits of data into *m* = 2*n* − 1 coefficients by flipping *t* coefficients.

Schonfeld and Winkler [21] found a way to hide data using more powerful error correction code (ECC). They used structured BCH code for data embedding. They showed two ways for computing position of the coefficients in the block to be modified. The first way is based on structured matrix as in matrix encoding; the second way uses the generator polynomial *g(x)*.

In this paper, a digital watermarking technique based on BCH error correction code in Discrete Cosine Transformation (DCT) domain is proposed. The overhead time is one of the disadvantages the BCH coding, regarding this challenge, this paper is also investigated lookup table method to decrease complexity of computation and speed up the process of finding the proper position to insert a watermark signal. Watermarked image robustness has been investigated against different kinds attacks and the simulation results indicate that the proposed algorithm outperforms the existing methods in terms of imperceptibility, robustness and security.

The rest of this paper is organized as follows. Section II introduces a basic definition of *t*-error-correction BCH coding scheme, BCH encoder and decoder. Lookup table method is introduced in Section III. In section IV, the proposed embedded and extraction algorithm is presented. Section V reports the experiments that are conducted for evaluating the proposed technique and presents the results. Finally, in the section VI, the conclusion is drawn.

## II. BCH CODE

The BCH code is named after Bose, Ray-Chaudhuri and Hocquenghem, who described methods in 1959 and 1960 for designing codes over *GF(2)* with a specified design distance. BCH codes that are constructed by using finite fields. BCH codes are a family of cyclic codes, with an algebraic structure that is useful in simplifying their encoding and decoding procedures. While operating under *GF(2$^m$)*, it has error correcting capability of *t*. The main parameters of BCH codes are summarized as following parameters [22, 23]:

- Block length: $n = 2^m - 1$
- Message length: *k*
- Maximum correctable error bits: *t*
- Number of information bits: $k \geq n - m*t$
- Minimum distance: $d_{min} \geq 2t + 1$

Clearly, this code is capable of correcting any combination of *t* or fewer errors in a block of $n = 2^m - 1$ digits. For a given codeword length *n*, only specific message length *k* is valid for a BCH code. This code is called a *t*-error-correcting BCH code. If a coding scheme generates a codeword of length *n* from a message of length *k*, the coding rate is $R = k/n$, $k \leq n$. A BCH code is generated by the polynomial g(x), the generator polynomial of BCH code is constructed by using minimal polynomials in GF(2$^m$) which is explained in [1]. Let $\alpha$ be a primitive element in GF(2$^m$). For any specified $m_0$ and $d_0$ the code generated by $g(X)$ is a BCH code, if $g(x)$ is the polynomial of lowest degree over GF(2$^m$) for which $\alpha^{m_0}, \alpha^{m_0+1}, \ldots, \alpha^{m_0+d_0-2}$ are roots. Let $\phi_i(x)$ be the minimal polynomials of $\alpha^i$ then generator polynomial of BCH code is computed as the least common multiple (LCM) among 2*t* minimal polynomials $\phi_i(x)$:

$$g(x) = LCM\{\phi_1(x), \phi_2(x),......, \phi_{2t}(x)\} \quad (1)$$

An (*n, k*) binary BCH code encodes *k*-bit messages into *n*-bit codewords. The *k*-bit message is the input of encoder and the BCH encoder generates (*n-k*)-bit parity. After encoding, the (*n-k*)-bit parity together with *k*-bit message generate a codeword as follows:

$$v(x) = c(x)x^{n-k} + \text{Re } m(v(x)x^{n-k})_{g(x)} \quad (2)$$

if channel is noisy, the received vector *r* can be expressed as follows:

$$r(x) = v(x) + e(x) \quad (3)$$

Upon receiving *r*, decoder must first determine whether there are transmission errors in *r*. If there is no error in the received data, the syndrome should be all zero and the decoding procedure is finished. Otherwise, the syndrome should be sent to error-location block in order to generate the error-locating polynomial. Then, the Chien search block is used to find out which bits are erroneous. Finally, the corrected message extracted. Generally, the decoding of BCH code has three main steps that are expressed as follows:

I. Computing the syndromes from the received codeword as follows:

$$S_i = r(\alpha^i) = c(\alpha^i) + e(\alpha^i) = e(\alpha^i) \quad (4)$$

$$= \sum_{j=0}^{n-1} e_i (\alpha^i)^j$$

II. Key equation solver, which determines the error locator polynomial $\sigma(X)$ through the BM (Berlekamp-Massey) algorithm as follows:

$$\sigma(X) = \sigma_0 + \sigma_1 X + \sigma_2 X^2 + ... + \sigma_v X^v \quad (5)$$

$$\sigma(X) = (1 + \beta_1 X)(1 + \beta_2 X)....(1 + \beta_v X)$$

III. Determining the error locating numbers by finding the roots of error locating polynomial (identifying the position of erroneous bit).

The block diagram of decoding process for a *t*-error correcting BCH code is illustrate in figure 1. For further information on BCH codes, we orientate readers towards references [22, 23].

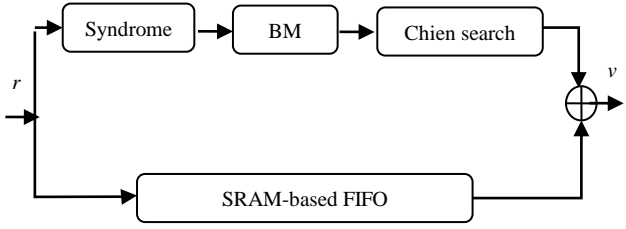

Fig. 1. BCH Block Diagram

Goal of coding theory is different to data hiding method even though both methods use the error correction codes [24]. In coding theory, syndrome equation $S = H_{k \times n} \cdot b^T$ in the first stage of decoder block is used to detect and correct error which is not expected in advance at the encoder block, whereas in data hiding, syndrome is used to choose proper coefficients to insert a watermark signal and decrease distortion. In the embedding process, by choosing proper $e(x)$, some of the host image bits are modified intentionally.

Let $r(x)$ and $v(x)$ denote watermarked image and host image respectively. $r(x)$ and $v(x)$ can be represented as the polynomial $r(x) = r_1 x + r_2 x^2 + ... + r_{n-1} x^{n-1}$ and $v(x) = v_1 x + v_2 x^2 + ... + v_{n-1} x^{n-1}$ over $GF(2^m)$. Then, watermark image $I$ is obtained as:

$$I = r.H^T \quad (6)$$

Where $H$ is the parity-check matrix that its entries are constructed by the primitive element in $GF(2^m)$ as follows:

$$H = \begin{bmatrix} 1 & \alpha & \alpha^2 & ... & \alpha^{n-1} \\ 1 & (\alpha)^2 & (\alpha^2)^2 & ... & (\alpha^2)^{n-1} \\ \vdots & & & ... & \vdots \\ \vdots & & & ... & \vdots \\ 1 & (\alpha^{2t}) & (\alpha^{2t})^2 & ... & (\alpha^{2t})^{n-1} \end{bmatrix} \quad (7)$$

Suppose that $v(x) = v_1 x + v_2 x^2 + ... + v_{n-1} x^{n-1}$ is host image and modify in the host image bits after insert watermark image results in the following watermarked image:

$$r(x) = r_1 x + r_2 x^2 + ... + r_{n-1} x^{n-1} \quad (8)$$

Let $e(x)$ be flip pattern. Then

$$r(x) = e(x) + v(x) \quad (9)$$

The first step of watermarking process is to compute the syndrome from the $r(x)$. The syndrome is a $2t$-tuple,

$$S = (S_1, S_2, ..., S_{2t}) = r.H^T \quad (10)$$

From (6) and (9) is obtained the following equation:

$$I = [v(x) + e(x)].H^T \quad (11)$$

$$I - v.H^T = e.H^T$$

From (10) and (11) the relationship between the syndrome and the flip pattern as follows, for further details, refer to reference [22].

$$S = I - v.H^T = e.H^T \quad (12)$$

Suppose that the flip pattern $e(x)$ has $v$ flip at locations $X^{j_1} + X^{j_2} + ... + X^{j_v}$ that is,

$$e(X) = X^{j_1} + X^{j_2} + ... + X^{j_v} \quad (13)$$

The power $j_1, j_2, ..., j_v$ tell us the location of host image bits that are modified in order to insert watermark image. Letting $B_k$ denotes $a^{jk}$, combining (12) and (13), the syndrome is obtained as follows:

$$S_1 = e_{j_1} \cdot \beta_1 + e_{j_2} \cdot \beta_2 + e_{j_3} \cdot \beta_3 + ... + e_{j_v} \cdot \beta_v$$
$$S_2 = e_{j_1} \cdot \beta_1^2 + e_{j_2} \cdot \beta_2^2 + e_{j_3} \cdot \beta_3^2 + ... + e_{j_v} \cdot \beta_v^2$$
$$\vdots \quad (14)$$
$$S_{2t} = e_{j_1} \cdot \beta_1^{2t} + e_{j_2} \cdot \beta_2^{2t} + e_{j_3} \cdot \beta_3^{2t} + ... + e_{j_v} \cdot \beta_v^{2t}$$

Notice that for binary codes, $e_{jk} = 1$ ($k = 1, 2, ..., k$). Equation (14) is further simplified to:

$$S_1 = \beta_1 + \beta_2 + \beta_3 + ... + \beta_{jv}$$
$$S_2 = \beta_1^2 + \beta_2^2 + \beta_3^2 + ... + \beta_{jv}^2 \quad (15)$$
$$S_3 = \beta_1^3 + \beta_2^3 + \beta_3^3 + ... + \beta_{jv}^3$$

Embedding process in watermarking is essentially done to find $B_k$'s. Once all $B_k$'s are known, the location of host image bits for insert watermark is also known. Equation (15) is a set of nonlinear function; solving it directly appears not to be a simple task. However, things will be easier if a polynomial σ(x) whose roots are the reciprocals of $B_k$'s:

$$\sigma(X) = (1 + \beta_1 X)(1 + \beta_2 X)...(1 + \beta_v X)$$
$$= \sigma_0 + \sigma_1 X + \sigma_2 X^2 + ... + \sigma_v X^v \quad (16)$$

The next step is to find the roots of error location polynomial. One simple yet effective method is to do an exhaustive search by using Chien search algorithm. Any root of σ(x) must be one of elements in Galois field. The Chien search examines if $\alpha^i$ is a root of σ(x) for $1 \leq i \leq n$ by substituting $\alpha^i$ into X of error location polynomial [22, 23].

Finding roots of error locator polynomial using Chien search algorithm leads to increase of latency in decoding process. In this paper an improved algorithm based the lookup table method for finding roots of polynomials over finite fields is utilized that is proposed by Zhao et al [25]. The effort of root finding can be significantly reduced by the use of lookup table method, it does not require exhaustive search. So, this makes possible fast embedding process to insert watermark signal using BCH codes. At the next section, a construction method of lookup table for quadratic and cubic polynomials is introduced based on the five lemmas which proposed by Zhao et al [25].

### III. LOOKUP TABLE METHOD

Consider pair of polynomial of degree 2 over finite field $GF(2^m)$ to be:

$$f(x) = x^2 + \sigma_1 x + \sigma_2$$
$$f(y) = y^2 + y + \sigma_2/\sigma_1^2 \qquad (17)$$

The method of Zhao states five lemmas as follows [25]:

***Lemma 1***: If $y_0$ is a root of $f(y)$, then $y_0 + 1$ is another root of $f(y)$.
***Lemma 2***: If $y_0$ is a root of $f(y)$, then $x_0 = \sigma_1 y_0$ is another root of $f(x)$.
***Lemma 3***: If $y_0$ is a root of $f(y)$, then $x_0 = \sigma_1 y_0 + \sigma_1$ is another root of $f(x)$.

Lookup table $q$ with size $n \times 1$ keeps the roots of any $y_0$ of the family of quadratic polynomials $f(y) = y^2 + y + i$, where $i \in [1; 2^m - 1]$ in $GF(2^m)$. Thus, the size of the lookup table $q$ is $(2^m - 1) \times 1$. The roots of polynomial $f(x)$ can be found in the lookup table $q$ in position $\sigma_2/\sigma_1^2$. If there are no roots for some corresponding position in the table is marked as -1 [7, 25].

For a cubic polynomial $f(x) = x^3 + \sigma_1 x^2 + \sigma_2 \cdot x + \sigma_3$ in $GF(2^m)$, there are two parameters such as $a = \sigma_1^2 + \sigma_2$ and $b = \sigma_1 \cdot \sigma_2 + \sigma_3$. If the polynomial $f(y) = y^3 + y + \dfrac{b}{a^{3/2}}$ be pair of $f(x)$, the relationship between roots of $f(x)$ and $f(y)$ can be described based on follow lemma:

***Lemma 4:*** If $y_j (j=1, 2, 3)$ is a root of $f(y)$, then $x_j = a^{1/2} y_j + \sigma_1$ is another root of $f(x)$.

The proof of these lemmas can be found in reference [25].

Similar to the quadratic polynomial, a lookup table $c$ with size $n \times 3$ keeps roots $y_j$ ($j = 1, 2, 3$) of the family of cubic polynomials $f_i(y) = y^3 + y + i$; If there are not three roots for some $f_i(y)$, the $i$th row of the table are marked as $-1$. The roots of the $f(x) = x^3 + \sigma_1 x^2 + \sigma_2 x + \sigma_3$ can be found in the lookup table $c$ in position $b/a^{3/2}$. For further calculations, the indexes of the rows which have 3 roots have to be stored in a special table $k$. The size of the special table $k$ is $D \times 1$, where $D$ is the number of rows which have 3 roots [7, 25].

### IV. PROPOSED WATERMARKING ALGORITHM

This section describes the process of data watermarking in digital image by using BCH coding. At the first step of the watermarking process, host image $v$ is divided into blocks $B$ of dimensions 8x8 pixels. Each block $B$ is transformed into frequency domain by DCT. The watermark bits are inserted in host image by using proposed image watermarking algorithm. The proposed image watermarking scheme is based on BCH($n$, $k$, $t$) where $n = 2^k – 1$, $m=1, 2, 3, ... , t = 2$ and able to insert $t.m$ bit watermark signal in the $n$ bit of host image. So proposed algorithm to insert watermark image $I$ ($|I|$ = t.m) in host image $v$ ($|v|$ = n) is designed as follows:

**Input :** watermark image $I$, blocks of host image, distortion maximum allowed $t =2, 3$.
**Output** : image watermarked $r$.
**Step 1** : Compute syndrome using equation (10).
**Step 2 :** If $S_1 = S_2 = 0$ there is not any $e$ in order to satisfy equation(12), omit current block, otherwise go to step 3.
**Step 3 :** If $S_2 + S_1^3 = 0$, watermark signal can be insert to a block of host image by modifying just one coefficient. So degree of polynomial $\sigma(x)$ is 1 such as $\sigma(X) = X + \sigma_1$ and its root is $\beta = S_1$. The position of the coefficient to be modified is $j_1 = \log(\beta_1)$, otherwise go to step 4.

**Step 4 :** If watermark signal inserting by modifying at most two coefficients be possible, The degree of polynomial $\sigma(x)$ is 2 such as $\sigma(X) = X^2 + \sigma_1 X + \sigma_2$. At the first, parameter $u = \dfrac{\sigma_2}{\sigma_1^2} = \dfrac{S_2 + S_1^3}{S_1^3}$ is calculated. The parameter $u$ is the index value in the lookup table $q$. The basic root $y_1 = q(u)$ can be obtained from the lookup table $q$ in row $u$. If value $y_1 \neq -1$, the roots of polynomial $\sigma(x)$ are computed such as $\beta_1 = S_1 \cdot y_1$, $\beta_2 = S_1 \cdot y_1 + S_1$ and the positions of the coefficient to be modified are $j_i = \log(\beta_i), i = 1,2$, otherwise go to step 5.

**Step 5 :** If watermark signal inserting by modifying at most three coefficients be possible, The degree of polynomial $\sigma(x)$ is $v = 3$ such as $\sigma(x) = x^3 + \sigma_1 x^2 + \sigma_2 \cdot x + \sigma_3$. At the first, parameter $o = \dfrac{b}{a^{3/2}}$ ( $a = \sigma_1^2 + \sigma_2$ and $b = \sigma_1 \cdot \sigma_2 + \sigma_3$ ) is calculated. The parameter $o$ is the index value in the lookup table $c$. The three basics roots $y_i = c(o,i), i = 1,2,3$, can be obtained from the lookup table $q$. The roots of polynomial $\sigma(x)$ are computed such as $\beta_i = \rho \cdot y_i + S_1, i = 1,2,3$ where $\rho = (\dfrac{s_1^3 + s_2}{o})^{1/3}$ and the positions of the coefficient to be modified are $j_i = \log(\beta_i), i = 1,2,3$.

**Step 6:** modifying in the host image bits $v$ after inserting watermark image results in the watermarked image as $r(x) = v(x) + e(x)$ and in order to embedding watermark signal, the proposed algorithm is repeated for each block.

In the extraction process distorted bits are retrieved by using BCH coding. Decoding process as explained in section II is applied to obtain secured watermark from each block. The BCH decoder can correct up to a certain number of errors. Thus, watermark image can be exactly extracted with no bit errors by this method.

## V. SIMULATION RESULTS

This section discusses results of several experiments conducted using the proposed algorithm. Simulations are carried out in MATLAB to test the proposed algorithm efficiency. It is notable to mention that Communications System Toolbox in MATLAB provides essential algorithms and functions for the implementation BCH code.

In these experiments, a secret image is concealed into different amount of host images using the proposed BCH-based watermarking scheme. The peak signal to noise ratio (*PSNR*) metric is measured in order to evaluate imperceptibility between original and attacked watermarked image. Its value can be defined as follows:

$$PSNR = 10\log_{10}\frac{255^2}{MSE} \quad (18)$$

Here MSE is called as the mean squared error between original and distorted image which is defined:

$$MSE = \frac{1}{MN}\sum_{i=0}^{M-1}\sum_{j=0}^{N-1}[x(i,j) - x'(i,j)]^2 \quad (19)$$

The better the watermarked image quality is, the higher the *PSNR* value will be. According to formula (18), the higher the *PSNR* value is, the lower the *MSE* value will be. Therefore, the better the watermarked image quality is, the lower the *MSE* value will be. In formula (18), *M* and *N* denote the length of the image, *x* is the pixel value of the original image, and *x'* is the pixel value of the watermarked image in position of (*i, j*). As pointed earlier, robustness is the resistance of an embedded watermark against different kinds attacks so, the normalized correlation coefficient (*NCC*) is used to evaluate the degree of between embedded watermark and extracted watermark from the attacked image which can be defined as follows:

$$NCC = \frac{\sum_{i=0}^{i-1}\sum_{j=0}^{j-1}W_{ij}W'_{ij}}{\sum_{i=0}^{i-1}\sum_{j=0}^{j-1}[W_{ij}]^2} \quad (20)$$

Where *i, j* define the size of the embedded watermark and *W* and *W'* define the original and extracted watermark bits, respectively. The value of *NCC* is between 0 and 1. And the bigger the value is, the better the watermark robustness is.

Figure 2 shows five test images were used in our experiments, namely barbara, watch, man, yacht, lena, and the size of each test image is 512x512. The cameraman image of size 256x256 is used as watermark image. In the first experiment we transformed each block of size 8x8 pixels of original image into DCT domain. Watermarked image imperceptibility, robustness, security and effectiveness BCH code have been investigated against different kinds attacks namely wiener filter, gaussian filter, rotation *2º*, JPEG compression and resize. Due to the limitation of paper space, we exhibit in figure 3 only extracted watermark for Lena image.

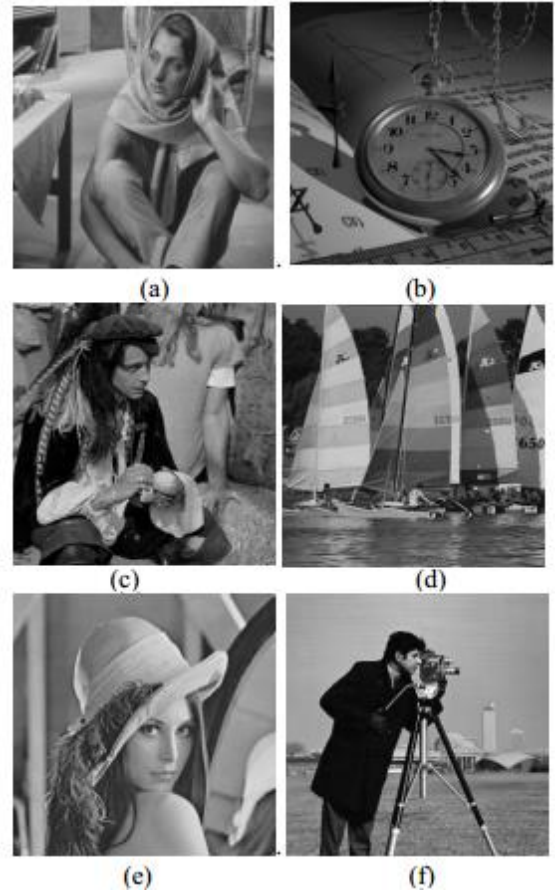

Fig. 2. Original Test Images and Watermark (a) Barbara (b) Watch (c) Man (d) Yacht (e) Lena (f) Watermark Cameraman

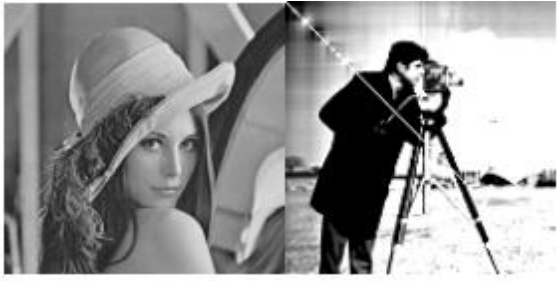

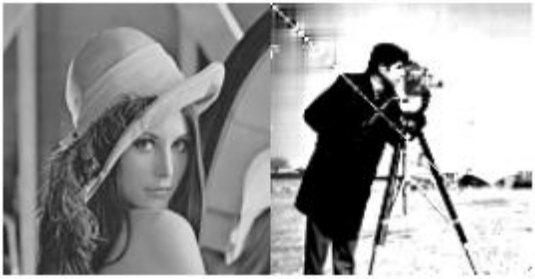

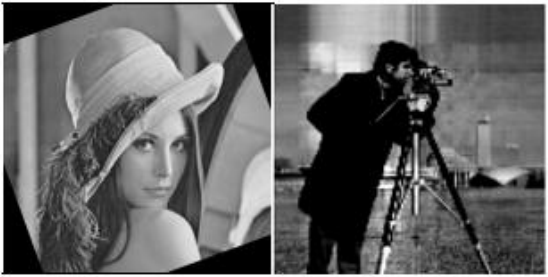

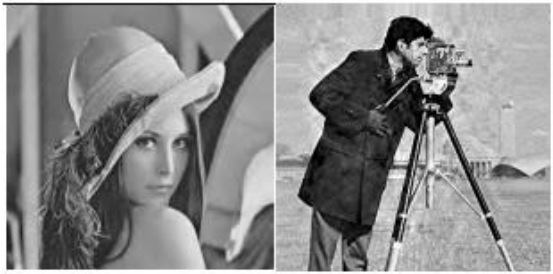

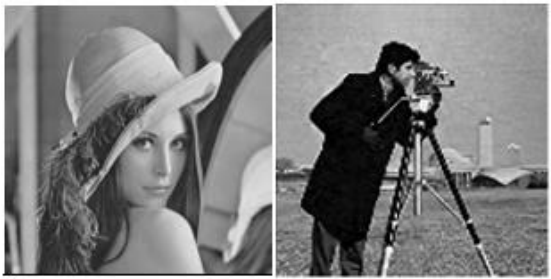

Fig. 3. Attacked image of watermarked Lena, and Corresponding extracted watermark after (a) 3x3 Median Filtering (b) Gauss noise (c) Rotation 2º (d) JPEG Compression (e) Resize 256→128→256

BCH coding parameters which are experimented in this study including: ECC rate *k/n*, maximum percent error correction capability of the ECC *t/n*, and *PSNR* value per codeword are given in Table 1. The relationship between *PSNR* of the watermarked image and the ECC rate is also shown in Table 1. It is observed that (31, 26), (31, 21) and (31, 16) BCH coding achieved the best results compared to (15, 11), (15, 7) and (15, 5) BCH coding which correct 1, 2 and 3 bits respectively.

There is a dependency between EEC rate and quality of the watermarked image. If ECC rate increases, redundancy is smaller, this means that distortion of original image with respect to embed watermark signal decreases, so watermarked image would be imperceptible. On the other hand, if ECC rate is low the quality of the image is low because the payload to insert is bigger. Generally ECC rates close to 1 have less redundancy and the payload is smaller, so the imperceptibility of the watermarked image (*PSNR*) is high. According to Table 1, if ECC rate be equal to 0.837, *PSNR* value is 42 dB and the watermarked image are imperceptible. For ECC rates lower than 0.516, the *PSNR* value and imperceptibility of the watermarked image reduces.

Table 1. Parameter BCH Coding in Proposed Watermarking Algorithm

| n | k | t | R = k/n | t/n | PSNR |
|---|---|---|---------|------|--------|
| 15 | 11 | 1 | 0.733 | 0.066 | 41.7 dB |
| 15 | 7 | 2 | 0.466 | 0.133 | 36.7 dB |
| 15 | 5 | 3 | 0.333 | 0.2 | 34.5 dB |
| 31 | 26 | 1 | 0.837 | 0.032 | 42.0 dB |
| 31 | 21 | 2 | 0.677 | 0.095 | 40.8 dB |
| 31 | 16 | 3 | 0.516 | 0.187 | 39.6 dB |

In this study, the parameters are used as the watermark strength $\alpha = 0.2$, The encoded message length *n = 31*, message length *k = 16* and error correction capability of the implemented BCH coding *t = 3* over $GF(2^5)$. For all attacks cases *PSNR* and *NCC* values are given in Table 2 and Table 3 respectively. It can be seen that the proposed algorithm is robust against most attacks. The values of *PSNR*s are between 35 and 42 dBs, and *NCC* values are close to 1 which confirm the desired robustness of proposed watermarking algorithm against different attacks.

In our experiments, we also test both methods: based only on the BCH-based watermarking scheme which uses Chien search method for finding position to insert watermark signal; and the BCH-based watermarking scheme with the proposed lookup table (LUT) method after applying different attacks on "Lena". Table 4 shows results are obtained using both method. The proposed method achieves high *PSNR* and *NCC* compared to classical method based on Chien search. It is evident that

Normalized Correlation Coefficient (*NCC*) values are greater than 0.90 and minimum *PSNR* is 38.68.

Table 2. Results of *PSNRS* Values for The Proposed Algorithm

| Attack | Name of Image | | | | |
|---|---|---|---|---|---|
| | Barbara | Watch | Man | Yacht | Lena |
| 3x3 *Wiener filter* | 41.50 | 40.62 | 41.72 | 40.89 | 41.69 |
| *Gaussian filter (Var=0.01)* | 40.52 | 40.52 | 40.63 | 39.14 | 40.59 |
| *Rotation 2º* | 39.82 | 39.69 | 39.56 | 37.63 | 39.43 |
| JPEG Compression | 40.81 | 39.50 | 39.45 | 39.49 | 39.68 |
| Resize 256→128→256 | 38.81 | 37.50 | 37.89 | 38.49 | 38.68 |

Table 3. Results of Normalized Correlation Coefficient (NCC) Values for the proposed algorithm

| Attack | Name of Image | | | | |
|---|---|---|---|---|---|
| | Barbara | Watch | Man | Yacht | Lena |
| 3x3 *Wiener filter* | 0.97 | 0.95 | 0.97 | 0.96 | 0.98 |
| *Gaussian filter (Var=0.01)* | 0.93 | 0.94 | 0.95 | 0.93 | 0.95 |
| *Rotation 2º* | 0.90 | 0.89 | 0.91 | 0.91 | 0.92 |
| JPEG Compression | 0.95 | 0.94 | 0.94 | 0.95 | 0.96 |
| Resize 256→128→256 | 0.96 | 0.96 | 0.95 | 0.96 | 0.95 |

Table 4. Comparison of proposed Algorithm with classical ones

| Attack | DCT_BCH Based LUT method | | DCT_BCH Based Chien search method | |
|---|---|---|---|---|
| | NCC | PSNR | NCC | PSNR |
| 3x3 *Wiener filter* | 0.98 | 41.69 | 0.76 | 27.05 |
| *Gaussian filter (Var=0.01)* | 0.95 | 40.59 | 0.61 | 20.05 |
| *Rotation 2º* | 0.92 | 39.43 | 0.74 | 17.69 |
| JPEG Compression | 0.96 | 39.68 | 0.73 | 29.69 |
| *Resize 256→128→256* | 0.95 | 38.68 | 28.49 | 32.68 |

Figure 4 shows the comparison of the visual quality with respect to the number of modified pixels between three algorithm: DCT-BCH-LUT, DCT-BCH-Chien search and DCT without BCH. This figure demonstrates two objectives. The first objective is to show image quality degradation with number of modified pixels to embed. The second objective is to show the proposed methods' excellent visual quality versus other methods results. The reason of proposed method has a better visual quality among other method is because the proposed watermarking algorithm based on BCH-LUT can find all possible positions to be modified and choose the best position for inserting watermark signal, which leads less distortion. Experimental results show the proposed algorithm has the higher operation in comparing with classical ones and detects accurately where the image has been modified, and it is able to resist large modification.

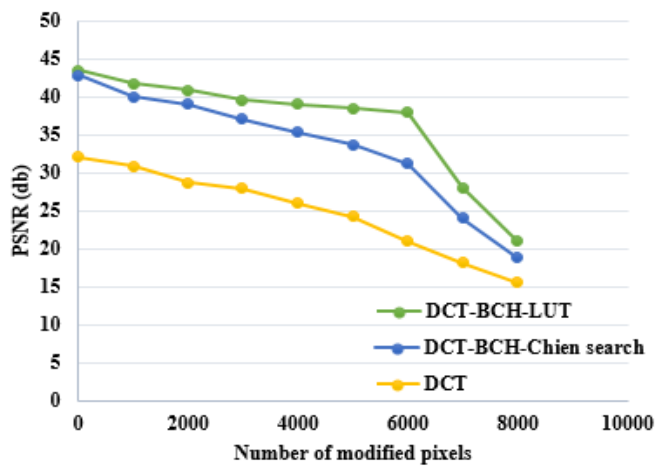

Fig. 3. Comparison of the visual quality of the proposed scheme with other schemes

## VI. CONCLUSIONS

In this paper an efficient digital watermarking algorithm based DCT and BCH error correction code is introduced. The proposed algorithm uses lookup table method in order to decrease complexity of computation and speed up the process of finding the proper position to insert a watermark signal. It can find all possible positions to be modified, and choose the best position for inserting watermark signal, which leads less distortion. The experimental results show that the proposed scheme, not only satisfies perceptual quality properties of watermarking, rather has more robustness in the operation of common image processing and geometric attacks in comparing the existing works.

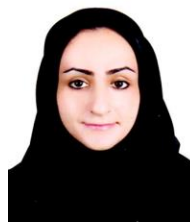

**Saeideh Nabipoour** received the B.Sc. degree in Computer Hardware Engineering from University of Guilan, Rasht, Iran and M.Sc. degree in Computer Architecture Engineering from University of Mohaghegh Ardabili, Ardabil, Iran, in 2012 and 2015, respectively. Her research interests include VLSI implementation of Error Correction Codes (BCH, Reed-Solomon), information theory and digital image processing.

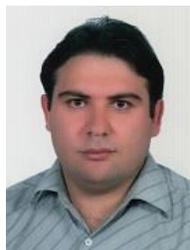

**Javad Javidan** received the B.Sc. and Ms.c. and Ph.D. degrees in Electrical Engineering from Sharif University of Technology, Tehran, in 2001, 2004 and 2011, respectively. Currently, he is an Assistant Professor at Mohaghegh Ardabili University, Ardabil, Iran. His research interests include power electronics and renewable energy, FACTS and HVDC.

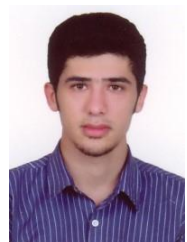

**Majid Khorrami** received the B.S. degree in Computer Software Engineering from the Poyandegan University of Chalus, Iran, in 2012. He received the M.Sc. degree in Computer Architecture Engineering from University of Mohaghegh Ardabili, Iran, in 2014. His current interests include digital image processing, machine vision and visualization.

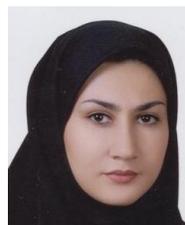

**Jila Azimzadeh** received the B.Sc. degree in Computer Hardware Engineering from University of Sadjad University of Mashad, Iran, and M.Sc. degree in Computer Architecture Engineering from University of Mohaghegh Ardabili, Ardabil, Iran, in 2012 and 2014, respectively. Her research interests include statistical modeling of images and videos, design of perceptual image and video quality assessment algorithms, and statistical data analysis.